\documentclass[adp,a4paper,fleqn%
]{w-art}
\usepackage{times,cite,w-thm}
\usepackage[]{graphicx}

\begin{document}
\DOIsuffix{theDOIsuffix}
\Volume{XX}
\Month{XX}
\Year{2009}
\pagespan{1}{}
\Receiveddate{XXXX}
\Reviseddate{XXXX}
\Accepteddate{XXXX}
\Dateposted{XXXX}


\title[Galactic halos in cosm. with LRSI]{Galactic halos in cosmology with long-range scalar DM interaction}

\author[W.~A. Hellwing]{Wojciech Hellwing\inst{1,}%
  \footnote{Corresponding author\quad E-mail:~\textsf{pchela@camk.edu.pl},
            Phone: +48\,22\,3296\,102,
            Fax: +48\,22\,8410\,046}}
\address[\inst{1}]{Nicolaus Copernicus Astronomical Center, ul. Bartycka 18, 00-716 Warsaw, Poland}

\def\hmpc{h^{-1}\,{\rm Mpc}}
\def\hkpc{h^{-1}\,{\rm kpc}}
\def\lcdm{\Lambda{\rm CDM}}

\begin{abstract}
Based on a set of cosmological N-body simulations we analyze properties of the dark matter haloes (DM) in a galaxy mass range ($10^{11} - 10^{13} h^{-1}M_{\odot}$) in modified $\lcdm$ cosmology with additional dynamically screened scalar interactions in DM sector. Our simulations show that scalar interactions support picture of the Island Universe. Rapid structure formation processes are shifted into higher redshifts resulting 
in a much smaller accretion and merging rates for galactic haloes at low redshifts. Finally, we present how this ``fifth'' force affects halo properties, 
like density profile, triaxiality, ellipticities and the spin parameter.
\end{abstract}
\maketitle                   

\section{\label{intro}Introduction}

In this paper we study the impact of the  long-range scalar interactions (LRSI) in the dark matter (DM) sector \cite{GP1,GP2} on some properties of the dark matter haloes. LRSI was proposed to overcome some difficulties that $\lcdm$ model faces at scales important to the galaxy formation. Large-scale clustering of LRSI cosmologies was studied in \cite{NGP,LRSI1,LRSI2}. Here we will study  DM haloes density profiles, mass accretion histories, shape and spin parameters. 

\section{\label{theory}Theory \& numerical application}

We model the LRSI in DM sector by introducing modifications to the Newtonian gravity.
Effective gravitational potential mimicking fifth force between two DM particles of mass $m$ is
\begin{equation}
\label{eqn:Yukawa}
\Phi({\bf r}) = -\,{Gm\over r}\left(1 + \beta\, e^{-x/r_s}\right) \; \;,
\end{equation}
Here $G$ is Newton's constant; ${\bf r}$ and ${\bf x} = {\bf r}/a(t)$ are, respectively, 
the particle separation vector in real and comoving coordinates. The $\beta$ parameter measures the relative strength of the scalar-interaction
compared to the usual gravitational interaction, while the $r_s$ parameters is Yukawa-like screening length and it measures the effective range of the scalar interactions.
Accordingly, the modified potential gives rise to a modified force law between DM particles. The new expression for the force law is
\begin{equation}
F_{DM} = -G{m^2\over r^2}\left[1+\beta\left(1+{r\over r_s}\right)e^{-r/r_s}\right]\;\; .
\label{eqn:force-law1}
\end{equation}

To study the structure formation and collapse of DM halos we use publicly available \verb#GADGET2# code by Volker Springel \cite{Gadget2}. \verb#GADGET2# is a hybrid N-body code that uses two different methods of calculating the particle forces. At large scales it uses a Particle-Mesh approach, an algorithm in which the Fast Fourier Transform method is used on a uniform lattice grid to solve 
the Poisson equation for gravitational potential. Then the potential is used to derive particles accelerations. We use Eqn. (\ref{eqn:Yukawa}) to modify the Green's functions in the $k$ space 
to obtain modified gravitational potentail. At small scales the code use Oct-Tree algorithm for obtaining the particle forces. We have implemented the Eqn. (\ref{eqn:force-law1}) in \verb#GADGET2# to obtain the modified force-law (see Ref. \cite{LRSI1} for the details). Finally to find haloes and calculate their properties from our simulations snapshots we have used the \textit{Amiga Halo Finder} (AHF), kindly provided to the public by A.~Knebe \& S.~Knollmann \cite{AHF}.

\section{\label{results}Results}

We have preformed a series of N-body simulations in boxes of width $25\hmpc$ containing $256^3$ DM particles. Therefore the mass resolution (mass of a single particle) in our numerical experiments was $m_p \sim 7.7\times 10^7 h^{-1}M_{\odot}$. Cosmology used in our simulation assume $\Omega_m = 0.3, \Omega_\Lambda = 0.7, \Omega_k = 0, h = 0.71$ and $\sigma_8 = 0.8$.

\begin{figure*}[b]
\includegraphics[width=0.37\linewidth, angle=-90]{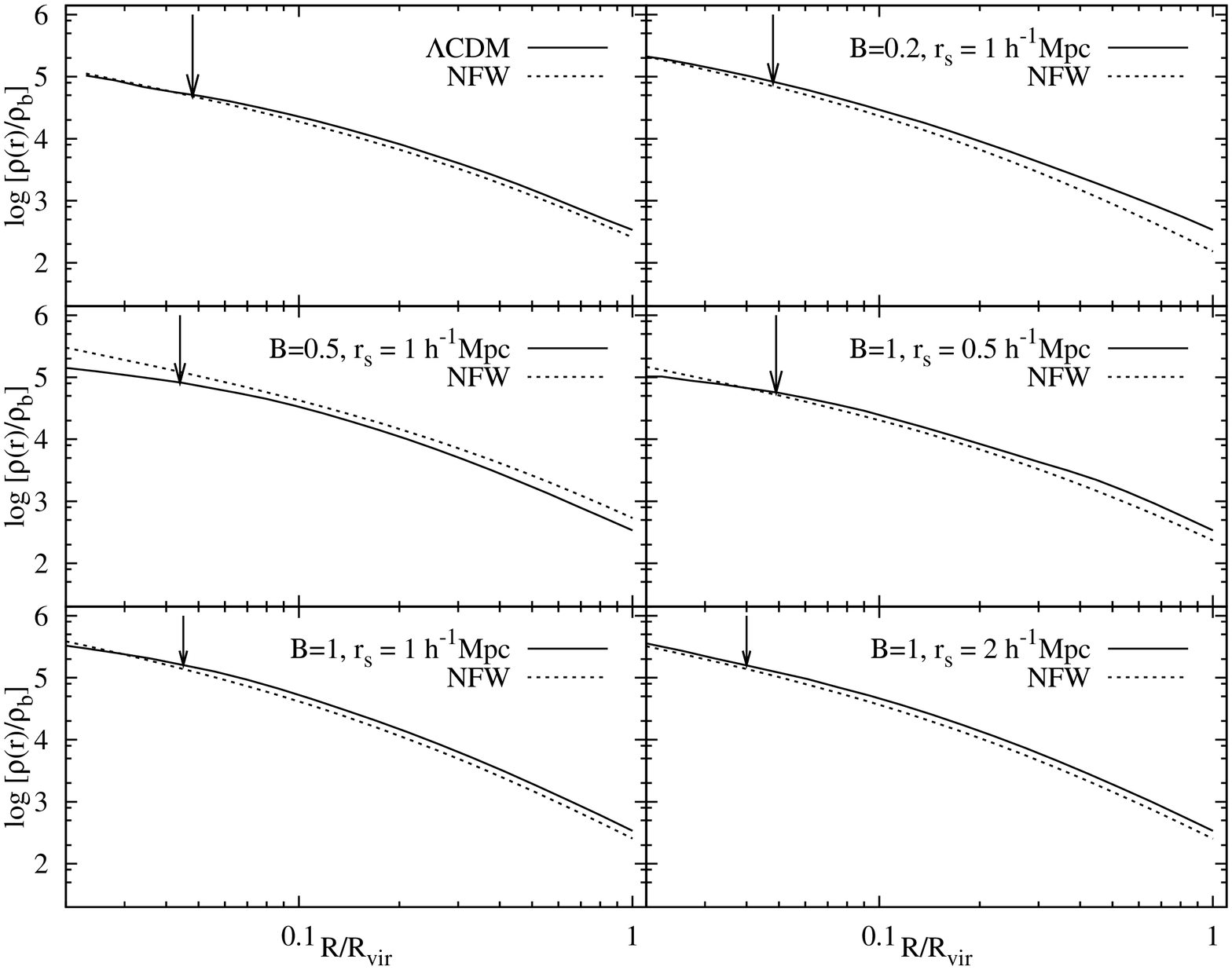}
\includegraphics[width=0.37\linewidth, angle=-90]{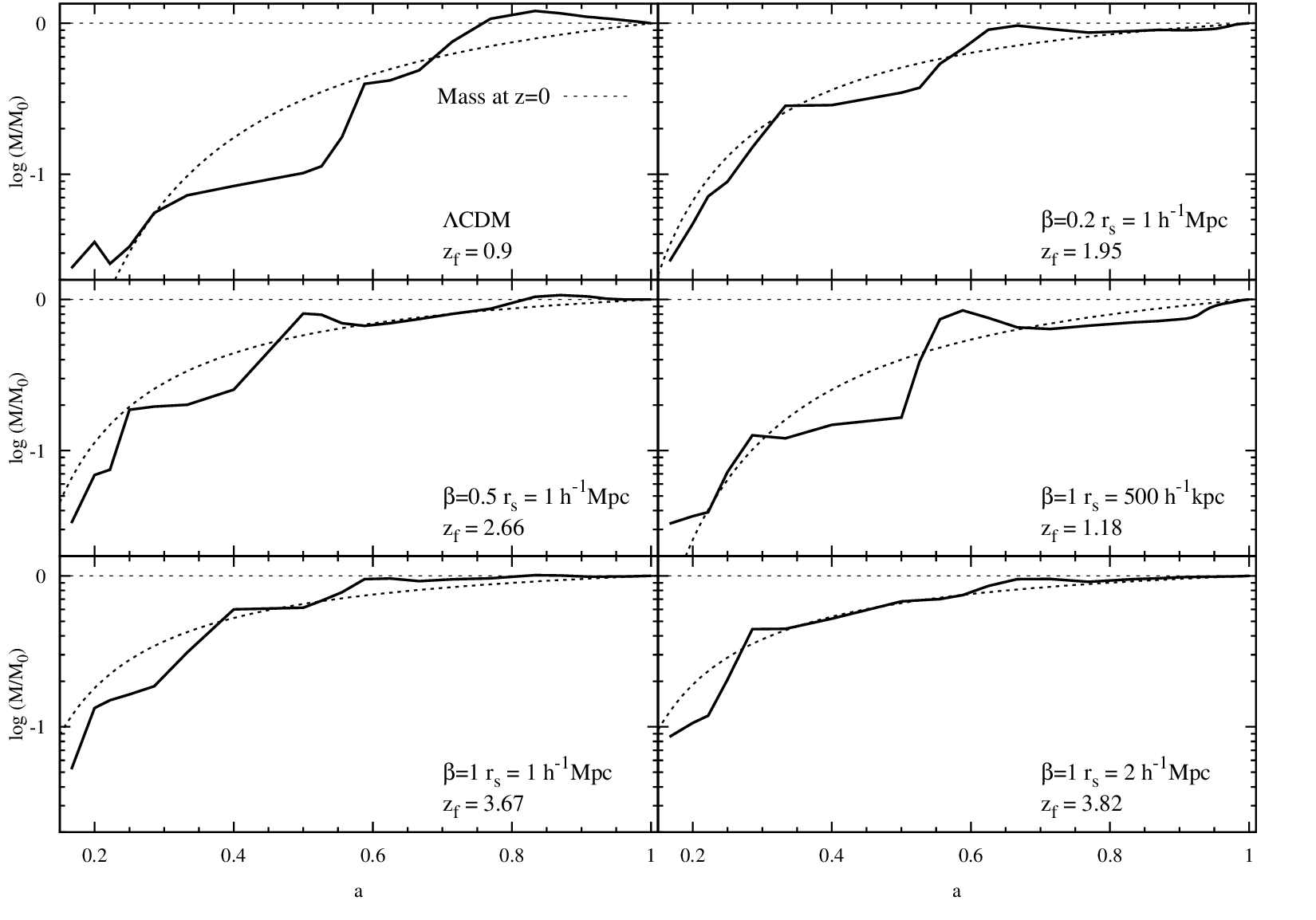}
\caption{\textit{On the left}: Density profiles of the cross-corelated haloes (\textit{solid lines}). \textit{Dotted lines} indicate NFW best fit. Arrows show the distance corresponding to the two times the force resolution $\approx 20\hkpc$. \textit{On the right}: Mass accretion histories of the cross-corelated haloes (\textit{solid lines}). \textit{Dotted lines} coresponds to the best fit of the equation (\ref{eqn:MAH}) to the data. \textit{Horizontal dashed line} indicates the final mass at $z=0$. In each panel at the bottom-right corner we present the formation redshift $z_f$ derived from best fit to the data. }
\label{fig:1}
\end{figure*}
To allow for a direct comparison between corresponding halo in different models, we identify a galactic halo in the $\Lambda$CDM run and then we cross-corelate (c-c) this halo with another halo found in the simulations with LRSI. As a criteria for matching halo we use: (a) - the same origin in space (in vicinity of $5\times r_{vir}$ at redshift of first identification); (b) - proximity of the final space occupation; (c) - adjacency of the velocity vector at $z=0$; (d) - the mass criteria (related halo in the LRSI model cannot excess five times the original $\Lambda$CDM halo mass). 

We begin by studying the shape of the density profiles of the c-c haloes. In the Fig. \ref{fig:1} on the left we show density profiles (solid lines) of the c-c haloes together with the best fits to the NFW profile \cite{NFW} (dotted line): $\rho_{NFW} = \rho_s/[(r/r_s)(1+r/r_s)^2]$,
where $\rho_s$ and $r_s$ are scaling density and scaling radius respectively. Arrows mark the distance from the center of the halo corresponding to the doubled force resolution $2\cdot\varepsilon\sim20 h^{-1}$kpc. We decided to not trust the results, below this distance, because they can be highly affected by two-body scattering. We note that for that particular halo NFW fit is not so good, especially for a halo in the model with $\beta = 0.2$, $r_s = 1h^{-1}$Mpc, and $\beta = 0.5$, $r_s = 1h^{-1}$Mpc. Nevertheless it is striking to depict, that the modified force-law and the modified potential do not change the universal shape of the halo density profile, known from the $\lcdm$ N-body simulations.

On the right side of the Fig. \ref{fig:1} we plot mass accretion histories (MAH's) of the c-c haloes (solid lines) versus the expansion factor $a$. Following the procedure presented in \cite{Wechsler2002} we fit MAH to the simple exponential law, given by:
\begin{equation}
\tilde{M}(a) = \exp\left[\alpha (1-a^{-1})\right],\qquad z_f = {2\over\alpha} - 1\;\;,
\label{eqn:MAH}
\end{equation}
where $\tilde{M}\equiv {M\over M_0}$, $M_0$ is the halo mass at $z=0$, $\alpha$ is the scaling parameter, $a$ is the cosmic scale factor, and $z_f$ is the formation redshift defined with the help of the $\alpha$ parameter.
We also plot best fits to MAH's for a single c-c halo. From these plots we see that c-c haloes in cosmologies with LRSI tend to have much higher formation redshifts $z_f$. Except cases with $\beta = 1$, $r_s = 500 h^{-1}$kpc and $\beta = 0.2$, $r_s = 1 h^{-1}$Mpc, c-c haloes in modified cosmologies have much quieter recent accretion histories then the $\Lambda$CDM case. Especially when we look at MAHs of c-c haloes with $\beta = 1$ and $r_s = 1$ or $2 h^{-1}$Mpc, we note that these haloes have not experienced any major merger since $a\sim 0.4\,\,(z = 1.5)$, and then they accreted mass steadily until present time.

\begin{table}
\caption{Summary of the properties of the cross-corelated halo in a different models at $z=0$. $M$ is the mass of halo $[10^{12}h^{-1}M_{\odot}]$, $\sigma_v$ is the 3D velocity dispersion $[km s^{-1}]$, $v_{max}$ is the maximum of the rotation curve $[km s^{-1}]$, $r_{vir}$ stands for the virial radii $[\hkpc]$, $T$ is the triaxialites parameters, $e_1$ and $e_2$ are ellipticities, $\lambda$ is the spin parameter and $z_f$ is the formation redshift as defined by the Eqn. (\ref{eqn:MAH}).}
\label{tab:1}
\begin{center}
\begin{tabular}{@{}lllllllllllllll@{}}
\hline
\hline
Model & $M$ & $\sigma_v$ & $v_{max}$ & $r_{vir}$ & $T$ & $e_1$ & $e_2$ & $\lambda$ & $z_f$ \\
\hline
\hline
$\lcdm$                         & 4.18 & 258 & 247  & 328 & 0.65 & 0.24 & 0.15  & 0.044 & 0.9\\
$\beta = 0.2$, $r_s = 1\hmpc$   & 5.12 & 303 & 266  & 351 & 0.93 & 0.23 & 0.21 & 0.047 & 1.95\\
$\beta = 0.5$, $r_s = 1\hmpc$   & 6.96 & 389 & 322  & 389 & 0.76 & 0.28 & 0.2 & 0.048 & 2.66\\
$\beta = 1.0$, $r_s = 0.5\hmpc$ & 6.86 & 454 & 312  & 387 & 0.41 & 0.37 & 0.13 & 0.103 & 1.18\\
$\beta = 1.0$, $r_s = 1\hmpc$   & 8.34 & 517 & 388  & 413 & 0.51 & 0.12 & 0.06 & 0.032 & 3.67\\
$\beta = 1.0$, $r_s = 2\hmpc$   & 12.15 & 572 & 426 & 468 & 0.3 & 0.13 & 0.037 & 0.053 & 3.82\\
\hline
\hline
\end{tabular}
\end{center}
\end{table}

Some properties of the cross-corelated haloes are given in the Table \ref{tab:1}. Note that corresponding halo in presence of the scalar interactions gets more massive at the final redshift.
The value of the velocity dispersion ($\sigma_v$) or the virial radius\footnote{We define virial radius as the radius at which the density of the halo drops to the value $\Delta\sim 340$ times the background density.} $r_v$ are also higher compared to the $\Lambda$CDM case. Now we define halo's shape and spin parameters as:
\[
\textrm{triaxiality:}\;\; T = {a^2-b^2\over a^2-c^2};\;\;\textrm{ellipticities:}\;\; e_1 = 1 - {c\over a}\;\; e_2 = 1 - {b\over a};\;\;\textrm{and spin:}\;\;\lambda={J\over\sqrt{2}M_{vir}v_{vir}r_{vir}}\;\;,
\label{eqn:params}
\]
where we adopted  definitions of the shape parameters from \cite{Franx_triax} and the spin parameter from \cite{Bullock2001}. Here $a>b>c$ are eigenvalues of the inertia tensor of a halo, $J$ is the total angular momentum of the halo and $M_{vir}, v_{vir}$ and $r_{vir}$ are its virial mass, velocity and radius respectively. Inspection of the properties collected in the Tab. \ref{tab:1} indicate that there is a big scatter between c-c haloes of their shape parameters, but we can note a weak tendency of higher $\lambda$ for stronger scalar interactions (except the case for $\beta = 1, r_s = 1\hmpc$). 

\begin{table}
\caption{Mean ellipticities, triaxiality and the spin parameter when averaging over haloes with $M\geq10^{11}h^{-1}M_{\odot}$.}
\label{tab:2}
\begin{center}
\begin{tabular}{@{}lllll@{}}
\hline
\hline
Model & $<e_1>$ & $<e_2>$ & $<T>$ & $<\lambda>$\\
\hline
\hline
$\lcdm$ & $0.25\pm0.09$ & $0.14\pm0.09$ & $0.58\pm0.22$ & $0.040\pm0.024$\\
$\beta = 0.2$, $r_s = 1\hmpc$   & $0.25\pm0.08$ & $0.15\pm0.08$ & $0.59\pm0.22$ & $0.044\pm0.028$\\
$\beta = 0.5$, $r_s = 1\hmpc$   & $0.25\pm0.08$ & $0.14\pm0.09$ & $0.57\pm0.23$ & $0.047\pm0.029$\\
$\beta = 1.0$, $r_s = 0.5\hmpc$ & $0.26\pm0.08$ & $0.16\pm0.09$ & $0.60\pm0.21$ & $0.052\pm0.03$\\
$\beta = 1.0$, $r_s = 1\hmpc$   & $0.25\pm0.09$ & $0.14\pm0.08$ & $0.57\pm0.21$ & $0.054\pm0.032$\\
$\beta = 1.0$, $r_s = 2\hmpc$   & $0.24\pm0.08$ & $0.13\pm0.08$ & $0.55\pm0.23$ & $0.053\pm0.04$\\
\hline
\hline
\end{tabular}
\end{center}
\end{table}
Now we turn to the statistical properties of the haloes formed in our simulations. We average three shape parameters and the spin parameter over all haloes found in simulation boxes which satisfy this mass criterion $M\geq 10^{11}h^{-1}M_{\odot}$. We present results in Table \ref{tab:2}. Clearly, the presence of the scalar interactions does not change shapes of the haloes. Triaxiality and ellipticities are converging to the values of the $\lcdm$ case within one sigma errors. On the other hand we can note, that halos in cosmology with the LRSI, tend to have higher spin parameters.

\section{\label{summary}Summary}

We have performed N-body simulations of structure formation in cosmologies with the additional long-range scalar interactions in the dark matter. We have studied the impact of such modification of the standard $\lcdm$ cosmological model on the c-c haloes and on the averaged statistical properties of the halo family. We conclude that the outer parts of the c-c halo density profile do not show any significant deviations from the $\lcdm$ halo density profile. We have to note though, that the c-c halos of our choice had their virial radii smaller then the screening length of scalar interactions models considered here. It will be interesting to study the inner most parts of the density profiles of cross-corelated haloes. To do that a much higher resolutions in forces and in masses are needed. We have also studied the MAH's of the c-c haloes. Here we see clear imprint of the LRSI on the mass accretion histories. Haloes in modified gravity have much higher formation redshifts, and they tend to show less mergers at small redshifts. This indicates that, in cosmology with the LRSI, small-scale structures like DM haloes form at earlier epochs and support the picture of galaxies being island universes at low redshifts. These results are encouraging in the view of apparent difficulties of reconciling the predictions from the $\lcdm$ cosmology galaxy formation simulations with astronomical observations (see Ref. \cite {NGP,LRSI1, 6puzzles} for detailed discussion). We have also measured averaged shape and spin parameters of the halo populations. We conclude that LRSI does not affect the shape of the haloes. However we see a clear signal in the spin parameter. DM haloes with LRSI have more angular momenta than their cousins in the $\lcdm$ case. The research presented here is just a first step in understanding how additional long-range interactions in the dark matter can affect various properties of the gravitationally bounded DM haloes. To study this matter in greater detail a more advanced study is needed. We have showed that modified gravity imprints on average halo spin and on particular halo mass accretion histories. Therefore it will be, in principle, possible to distinguish between the $\lcdm$ and LRSI models using data from future detailed galaxy surveys.

 
\providecommand{\WileyBibTextsc}{}
\let\textsc\WileyBibTextsc
\providecommand{\othercit}{}
\providecommand{\jr}[1]{#1}
\providecommand{\etal}{~et~al.}

\end{document}